# Second-Harmonic Generation Enhancement by Film-Coupled Plasmonic Nanopatch Antennas With Integrated Quantum Emitters

*Bryson Krause,[1] Dhananjay Mishra,[2] Jiyang Chen,[1] Christos Argyropoulos[2] and Thang Hoang[1,]\**

[1]Department of Physics and Material Science, University of Memphis, Memphis, TN 38152

[2]Department of Electrical and Computer Engineering, University of Nebraska-Lincoln, Lincoln, NE, 68588

\*E-mail: tbhoang@memphis.edu

Plasmonic nanocavities have proven to be a powerful optical platform for confining light at a length scale much smaller than the diffraction limit. Enhanced electromagnetic fields within the cavity mode volume enable multiple significant effects that lead to applications in both the linear and nonlinear optical regimes. In this work, we demonstrate enhanced second harmonic generation from individual plasmonic nanopatch antennas which are formed by separating silver nanocubes from a smooth gold film using a sub-10 nm zinc oxide spacer layer. When the nanopatch antennas are excited at their fundamental plasmon frequency, a $10^4$-fold increase in the intensity of the second harmonic generation wave is observed. Moreover, by integrating quantum emitters that have an absorption energy at the fundamental frequency, a second order nonlinear exciton-polariton strong coupling response is observed with a Rabi splitting energy of 19 meV. The nonlinear frequency conversion using nanopatch antennas thus provides an excellent platform for nonlinear control of the light-matter interactions in both weak and strong coupling regimes which will have a great potential for applications in optical engineering and information processing.

**Keywords**

nonlinear optics, second harmonic generation, plasmonics, nanocavity, patch antenna

# 1. Introduction

Many nonlinear up-conversion processes such as second and third harmonic generation and wave mixing of low frequency light to higher frequencies have been investigated extensively in the past few decades.[1-8] These nonlinear processes can have a wide range of applications from laser technology to biomedical imaging and sensing.[9-17] Traditionally, nonlinear processes such as second and third harmonic generations are demonstrated in bulk crystal materials with given specific symmetries, such as noncentrosymmetric or centrosymmetric media, respectively. Additionally, for efficient nonlinear optical processes, usually perfect phase-matching conditions are desired. Therefore, in practice, for efficient nonlinear frequency conversion, macroscopic birefringent crystals are often used as the nonlinear medium, and a high intensity input field is also required. Recent advances in the field of plasmonics have shown that the nonlinear generation efficiency can be enhanced significantly by using nanoscale metallic structures.[18-22] A plasmonic nanostructure can localize strong electromagnetic fields to a volume much smaller than the diffraction limit, resulting in a greatly enhanced electromagnetic field intensity which is essential for nonlinear optical processes. Further, due to the ability of these structures to generate enhanced fields in the nanoscale region, perfect phase-matching is usually not required in plasmonic nanostructures.[23] Indeed, several previous nonlinear frequency conversion processes have been realized by using plasmonic nanocavities ranging from individual nanocavities such as single plasmonic nanoparticles[24-26] to arrays of nanoparticles (or metasurfaces).[1, 7, 27-31] In practice, gold (Au) is widely used to fabricate plasmonic nanocavities. However, Au is a centrosymmetric material that in principle prohibits bulk second harmonic generation (SHG) while allowing third-order nonlinear processes. At the nanoscale, where the surface-to-volume ratio is large (on the order of $10^8$ $m^{-1}$), the symmetry of the Au crystal is broken at the surface, so both second and third nonlinear processes are allowed. Also, a general approach to enhance optical nonlinear processes by using plasmonic nanostructures is to overlap the fundamental input wave with the plasmonic resonance, which eventually helps increase the absorption rate of the input field.[1, 7, 9, 25, 28, 29, 32] Several other attempts have also tried to match resonant modes of plasmonic structures with both the fundamental input frequency and the nonlinear signal frequencies. Further, improved phase-matching conditions for highly efficient nonlinear optical processes have also been realized by using plasmonic nanocavities. These efforts of using plasmonic platforms to significantly enhance various nonlinear optical processes, such as second and third harmonic generation, four-wave-mixing or sum of frequency generation, have sparked a tremendous interest in research and have motivated growth in potential applications. Recent articles[28, 33, 34] have discussed ways to enhance SHG by introducing mode-matching conditions at the fundamental and second harmonic frequencies of nanocube-film coupled nanopatch antennas (NPAs). Nevertheless, it appears that there is still much more to explore in order to achieve highly efficient SHG, for instance, by matching the localized surface plasmon resonance of the stand-alone nanocube or one of the higher-order Fabry-Pérot (FP) modes of the gap mode to appear exactly at the second harmonic frequency of the fundamental mode. In addition, there is a lack of discussion in the literature regarding conditions for phase-matching between fundamental and second harmonic modes for efficient SHG conversion by using NPAs. The effect of strong coupling in the nonlinear regime[35] has also not been investigated with these emerging NPA platforms that can achieve extremely strong light-matter interactions in the nanoscale.

In this work, we demonstrate that the SHG is significantly enhanced by employing a gap-mode plasmonic nanocavity based on the NPA design. The gap-mode plasmon resonance frequency of the film-coupled nanocube NPA, which describes an enhanced photonic density of states in the sub-10 nm gap volume sandwiched between the nanocube and film, is designed to overlap with the fundamental input frequency. Furthermore, the SHG frequency is coincided with the resonance frequency of stand-alone nanocubes which can further help enhance the conversion efficiency. Compared to other types of plasmonic platforms that have been demonstrated to enhance the SHG, NPAs offer the highest field enhancement through the formation of FP or gap-plasmon resonances leading to hot spots within the thin gap layer.[36-40] In our study, by measuring the SHG signals from ensembles of NPAs as well as from individual NPAs, we observed a SHG enhancement factor of $10^4$ for an ensemble of NPA structures as compared to the bare Au film. Furthermore, by integrating organic dye molecules, which have a strong absorption band at the fundamental resonance of the NPA, with the cavity we observed second harmonic polaritons with a Rabi splitting of 19 meV. This interesting effect is due to the strong exciton-plasmon coupling[41] at the fundamental resonance of the NPA that is substantially enhanced and being transferred coherently to the second harmonic frequency through the upconversion process. Note that nonlinearities have been found to increase the sensitivity of plasmonic sensors[42] but very limited work exists to demonstrate nonlinear-mediated strong coupling which is clearly shown in our current work.

## 2. Methods

The plasmonic NPA consists of 100-nm silver (Ag) nanocubes situated on a smooth 100 nm-thick Au film. A ZnO layer separation between the nanocube and Au film is deposited via atomic layer deposition (ALD). The fabrication procedure of the NPAs is briefly described as follows: first, a Au film is deposited by an electron beam evaporator on a polished silicon wafer and the Au is subsequently removed from the silicon substrate and adhered to a glass substrate by means of the template stripping technique.[43] In this way the smoothness of the Au film is defined by the polished silicon surface which is atomically flat. A ZnO film (1 – 5 nm) is then deposited on top of the Au film by the ALD method. The thickness of the ZnO layer is confirmed by an atomic force microscopic measurements (see Supporting Information, Figure S1). Alternatively, instead of the ZnO layer, the Au film can also be coated with a layer of polyelectrolyte (PE) polymer spacer through a dip coating technique.[44] Subsequently, Ag nanocubes (purchased from Nanocomposix) are randomly suspended on top of the spacer layer to form NPAs. Figure 1(a) displays a scanning electron microscopic image of a finished sample which clearly shows well-separated nanocubes on the surface. Statistically we observed that deposited nanocubes cover an approximate 5% total area of the sample.

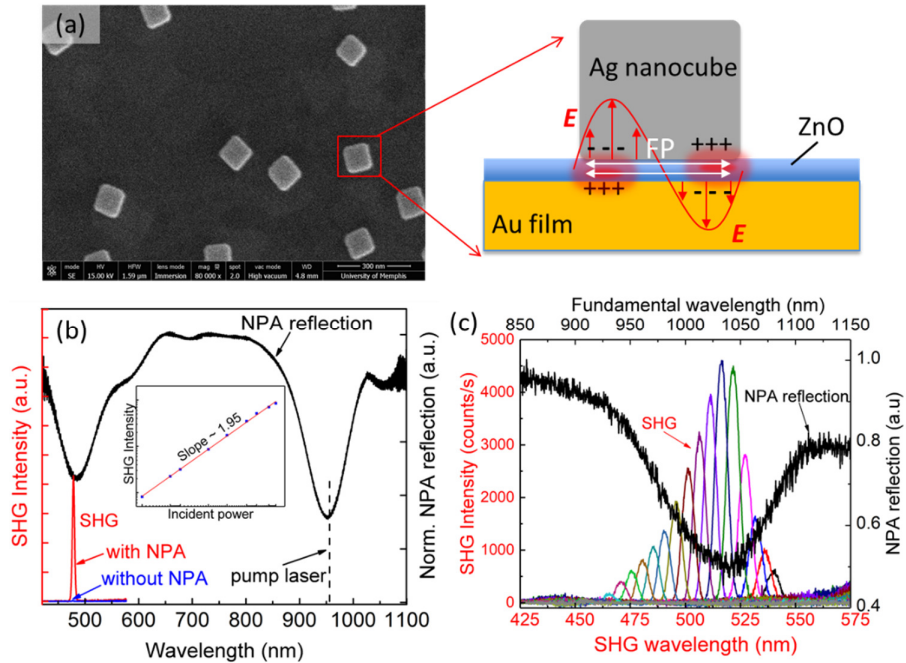

**Figure 1. Plasmonic NPAs for SHG enhancement.** (**a**) SEM image of a NPA sample. The schematic shows the sample structure and sketch of the optical mode at the fundamental resonance. (b) Representative reflectance spectrum (black) of NPAs composed of 100 nm nanocubes separated from a Au film by a 3 nm ZnO layer. The red curve shows the pump laser while the green and blue curves show the SHG signal from samples with and without NPAs. The inset shows the power dependent SHG intensity. (c) Excitation wavelength dependent SHG signal (left/bottom axes) and the reflectance spectrum at the fundamental frequency (top/right axes).

In the optical measurement procedure, the fundamental plasmonic resonant mode of the sample is first characterized by measuring the white light reflection spectrum from an ensemble of NPAs. Figure 1(b) shows such a reflection spectrum for NPAs fabricated with a 3 nm ZnO gap thickness. A reflection dip at approximately 960 nm indicates a strong absorption band, and the localized gap-plasmon resonance occurs at this wavelength. Once the fundamental frequency of the NPAs is determined, a tunable laser is tuned into resonance with the mode of the NPA and the SHG signal is recorded (Figure 1(b)). Figure 1(b) also displays the SHG from a control sample, i.e., a bare 100-nm thick Au film without a ZnO layer or nanocubes. It is clear that the NPAs have significantly enhanced the SHG signal. The inset of Figure 1(b) shows the laser excitation power dependence on the SHG signal. The integrated peak intensity is plotted against the laser pump power and the slope of the fitted dependence line is determined to be 1.95 ± 0.02, which is very close to 2, verifying a second power order dependence. It is also important to note that there is another plasmon resonance in the NPA system at approximately 485 nm, which is related to the resonance of stand-alone nanocubes.[45] As a result, the SHG frequency at 480 nm is within the full-width-at-half-maximum of the nanocube's stand-alone resonance which further enhances the SHG output intensity. We further verify the role of the NPA plasmonic gap-mode resonance in the SHG process by tuning the wavelength of the input laser across the resonance peak, as shown in Figure 1(c) for an ensemble of NPAs that have a 1 nm ZnO spacer gap layer. As expected, while the input laser intensity is unchanged, the SHG peak intensity is strongest when the fundamental wavelength is at the minimum of the reflectance curve. In other words, at the minimum of the reflection of the NPAs, the absorption of the input laser is maximum leading to the most enhanced SHG process.

## 3. Results and Discussion

Figure 2(a) compares SHG signal intensities for different substrates and NPAs, which have a fundamental resonance at 960 nm, using the same excitation laser wavelength and intensity. First, for the same 3-nm layer of ZnO on glass, the SHG intensity is unnoticeable. For the 3-nm ZnO layer on a Au film (100-nm thick), the SHG signal is very weak and identical to that for SHG from the bare Au film. However, when the Ag nanocubes are included, SHG is substantially enhanced. These measurements again provide strong evidence that the plasmonic gap mode has a significant role in boosting the SHG process efficiency.

In order to quantify a fair enhancement factor we compared the SHG intensities from NPAs and a bare Au film, while taking into account a normalization factor for the area covered by the nanocubes (~ 5%) and the laser excitation spot size (~ 5 µm). An enhancement factor of 10,700 is extracted for an ensemble of NPAs. While our experimental setup does not allow us to determine the SHG conversion efficiency, the measured SHG enhancement factor of our work is comparable to that of periodic arrays of NPAs as reported by Zeng et al.[28]. Nevertheless, in our NPA design, a ZnO spacer gap layer is used in contrast to a polymer shell which can potentially be damaged by an intense input laser that is needed to achieve high SHG efficiencies. Figure 2(b) compares SHG from NPAs with two different spacer materials, namely, ZnO and PE polymer, while using the same excitation conditions. It is clear that SHG effciency is strongly dependent on the spacer gap. This is in agreement with the fact that the second-order susceptibility of ZnO (~ 10.2 pm/V)[46, 47] is higher than that of PE polymer (~6.5 pm/V)[48] so the SHG from NPAs with a ZnO gap is higher than the SHG from NPAs with a PE gap. However, it is also important to note that the metal-dielectric interfaces may play a role besides the value of the second-order susceptibilities. Previous investigations using gap-mode plasmonic platforms to enhance SHG efficiency have led to several different conclusions regarding the origin of the upconversion process. Some have focused on the use of nonlinear materials within the nanogap,[1, 7] while others have argued that the metal/dielectric interface is also responsible for the enhanced SHG.[49]

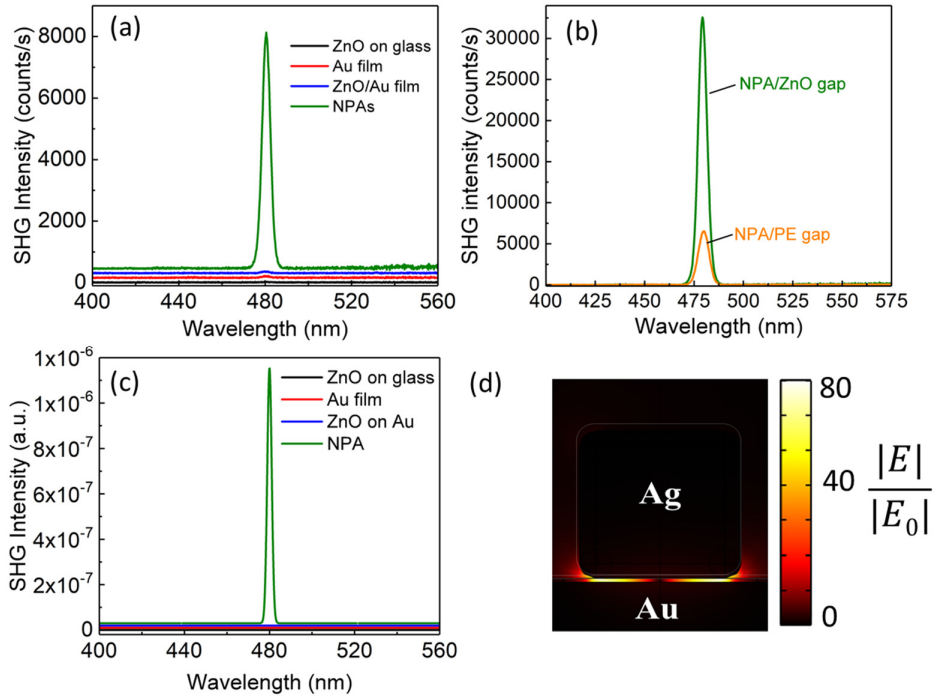

**Figure 2. Substrate dependent SHG intensities.** (a) Comparison of measured SHG spectra from NPA, ZnO, Au and ZnO/Au. The vertical axis is slightly offset for clarity. (b) Comparison of the SHG spectra from NPAs with ZnO and PE spacer gap. (c) Computational study of the SHG spectra from NPA using a ZnO spacer gap, ZnO, Au and ZnO/Au. (d) Computed field distribution at the fundamental frequency of NPAs by using ZnO as the spacer layer.

The SHG is theoretically investigated by using COMSOL Multiphysics nonlinear simulations to model the three different substrates (3nm ZnO on glass, 3nm ZnO on Au film, 100nm Au film) and the proposed NPA structure. Figure 2(c) verifies that the SHG is significantly enhanced only by the NPA structure which is in excellent agreement with Figure 2(a). We note that the results presented in Figure 2(c) were obtained by multiplying the continuous wave simulation results obtained in the frequency domain with a frequency-domain Gaussian profile that imitates the envelope of the pulsed laser illumination in our experiment. The field enhancement induced along the structure by the incident wave is calculated by the maximum ratio $E/E_0$, where $E_0$ represents the incident electric field amplitude of the input wave and $E$ is the induced maximum electric field in the nanogap. The result at the fundamental resonance is demonstrated in Figure 2(d). Further details about the linear and nonlinear simulations are provided in the supplementary information.

Next, we present the SHG enhancement results from measurements of single NPAs. Several previous works have reported the SHG enhancement from an ensemble of NPAs but none has shown nonlinear-scattering measurements of single NPAs. Indeed, the SHG enhancement measurements from a single NPA is somewhat technically challenging because of the relatively weak nonlinear signal as well as weak light scattering at the fundamental frequency. In our approach, we first identify individual NPAs by imaging SHG from a diluted sample. Once an NPA is identified, it is spatially filtered by a pinhole at an image plane and is spectrally characterized in both the SHG intensity and dark field scattering.

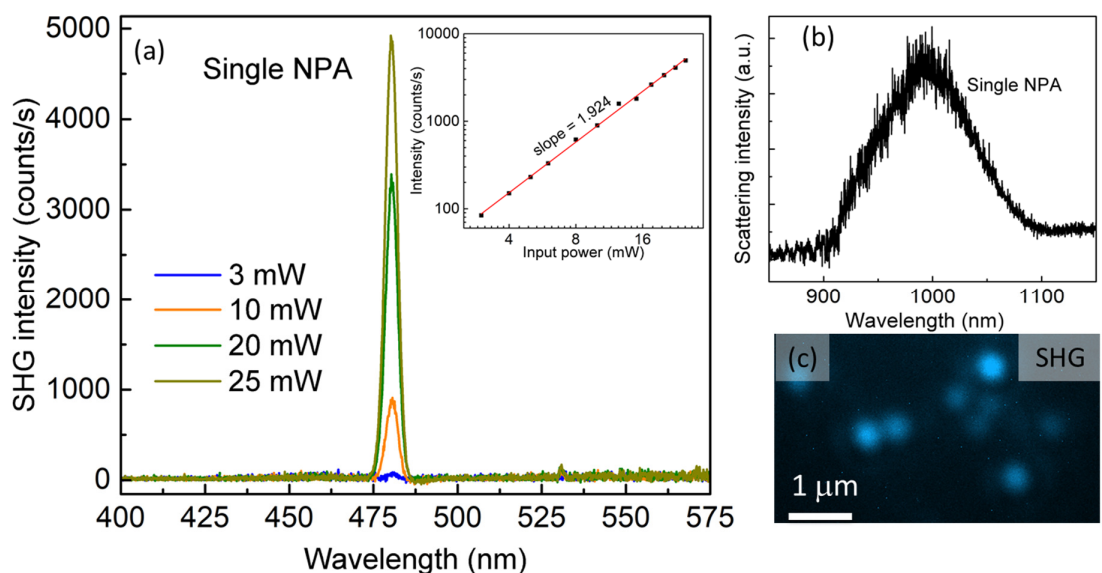

**Figure 3. SHG of single NPAs.** (a) Power dependent SHG spectra from a single NPA. The inset shows the excitation power dependent SHG intensity. (b) Dark-field linear scattering spectrum of the NPA measured in (a). (c) Camera image of SHG signal from individual NPAs.

Figure 3(a) shows the measured SHG spectra from a single NPA by using various excitation powers. The dark-field linear scattering spectrum of this particular NPA is shown in Figure 3(b). By tuning the laser fundamental wavelength to near the maximum of the scattering spectra, the SHG is recorded and the input-output relationship between the fundamental laser and the SHG signal is shown in the inset of Figure 3(a). Once again, the power dependent slope (~ 1.924) is very close to 2 indicating the second order nonlinear process nature. Figure 3(c) shows the SHG camera image of several isolated NPAs which are spectrally filtered by a bandpass filter centered at 480 nm. By comparing the integrated SHG peak intensity from a single NPA with a bare Au film and taking into account the normalization factor due to the laser excitation spot size and the size of the nanocube (100 nm), we extract an enhancement factor of 2,070 for this particular isolated NPA. The extracted enhanced value for this particular NPA is different from the value obtained for an ensemble of NPAs as described above, which is attributed to the unavoidable variation among NPAs and their gap thicknesses (as a result of roughness of the ZnO spacer layer, for example).

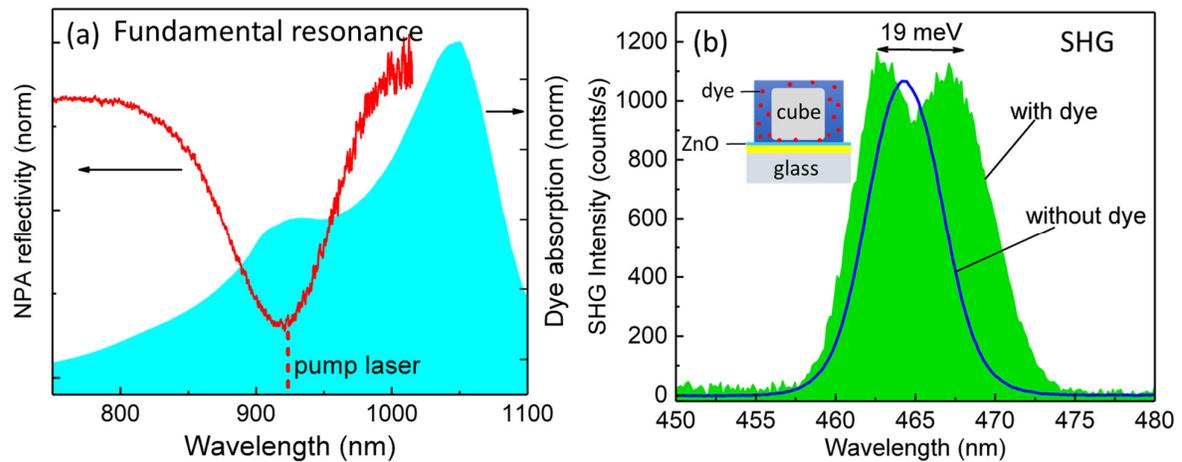

**Figure 4. SH exciton-polaritons and strong coupling in the nonlinear regime.** (a) Linear NPA reflection (without dye, red) and the IR-1048 dye absorption (cyan, without NPA). (b) SHG nonlinear signal with and without dye. Substantial resonance splitting in SHG is obtained only in the case of strong coupling (with dye curve). The inset shows a schematic of the dye/NPA structure.

Our next step in attempting to investigate more complicated nonlinear light-matter interactions, such as nonlinear strong coupling response, is to integrate quantum emitters (IR-1048 dye molecules suspended in DMSO (Dimethyl Sulfoxide)) with NPAs. Because the IR-1048 has a strong absorption band in the 900-1050 nm range, the NPAs are designed to have gap-mode resonance in this frequency range. Figure 4(a) presents the linear characterizations of NPA reflection (without dye, red curve) and IR-1048 dye absorption (cyan, without NPA, cyan curve). To improve the detection efficiency of our optical system, we have chosen to work with the higher dye absorption band near 930 nm. Therefore, Figure 4(a) shows a clear spectral overlap between the NPA fundamental mode and the dye absorption band at approximately 940 nm, which is important for exciton-plasmon coupling.[41] Figure 4(b) shows SHG measurements from the same NPA sample in two cases, with and without dye molecules, by using the same input pump laser power of 6 mW. Interestingly, when the IR-1048 dye is included, the SHG peak broadens and splits to two distinguished peaks, which are attributed to the upper and lower nonlinear second-order polaritonic states. Hence, the strong exciton-plasmon coupling in the presence of the NPAs leads to the formation of polaritons, a hybrid light/matter state, at the fundamental frequency, and these are subsequently upconverted to SH nonlinear polaritons with a pronounced Rabi splitting $\Delta E = 19$ meV. It should be noted that the SH nonlinear polaritons are observed with a relatively high dye concentration density (~ 65 mM) and tend to be unstable under higher excitation input power mainly due to the dye quenching effect. In an earlier theoretical work,[30] Drobnyh and Sukharev demonstrated a Rabi splitting up to 68 meV at the SH frequency by incorporating emitters that have an absorption band at the fundamental frequency of nanohole arrays. A recent work by Li et. al.[35] has experimentally demonstrated a large Rabi splitting at the SH frequency of a strongly coupled plasmonic nanorod and $WSe_2$ monolayer. We believe that for an optical platform that has a narrower resonance bandwidth (or higher quality factor) than the currently employed NPAs, a wider energy splitting and more pronounced nonlinear polaritonic states can be achieved. We would like to note that in principle one could also observe the exciton-plasmon coupling at the fundamental frequency by observing an energy splitting in the reflectance spectrum of an NPA/dye hybrid structure or in the scattering response of a single NPA.[41] However, the relatively broad resonance band of the NPAs (~ 200 meV) is approximately twenty times broader than the expected energy splitting (should be approximately equal to $\Delta E/2$) at the fundamental frequency. It is therefore very challenging to observe such a splitting in the white light reflectance spectrum (Figure 4(a) at the fundamental frequency. Indeed, we have recorded the reflection spectrum of the pump laser itself as it is scanned through the resonance of the NPAs in order to observe an energy splitting at the fundamental frequency. Such a measurement result is presented in the Supporting Information (Figure S2), where an energy splitting of approximately 9 meV (~$\Delta E/2$) is in fact observed. Further, the energy splitting value will be dependent on the input pump laser intensity as well as the emitter density. Therefore, for future studies of nonlinear polaritonic states robust quantum nanomaterials such colloidal quantum dots (for example PbS quantum dots) or 2D materials (such as black phosphorous) can be used.

## 4. Conclusion

In conclusion, we have investigated the enhanced SHG by using gap-plasmon mode NPAs. The nanocube-Au film hybrid structure enhances SHG by up to four orders of magnitude. More interestingly, by integrating quantum emitters within the NPA system we demonstrate the SHG nonlinear response of the exciton-polariton with a large Rabi splitting. Our results pave the way for optical designs that could support nonlinear polaritonic states and even long-range nonlinear polariton condensates in NPA arrays forming metasurfaces, which are expected to be very important for quantum information science.

**Acknowledgements**

This work is supported by the National Science Foundation (NSF), DMR-1709612 (BK, TH and CA), by the Ralph E. Powe Junior Faculty Enhancement Awards from Oak Ridge Associated Universities (to TH). TH acknowledges the National Institute of Health (NIH Grant # R15 CA238890-01A1) for its support. CA acknowledges partially support from the Office of Naval Research Young Investigator Program (ONR-YIP) (Grant No. N00014-19-1-2384) and NSF/EPSCoR RII Track-1: Emergent Quantum Materials and Technologies (EQUATE) under (Grant No. OIA-2044049). The authors would like to thank Ms. Martina Rodriguez Sala for her assistance with the IR-1048 dye absorption measurement.

Supporting Information

# Second-Harmonic Generation Enhancement by Film-Coupled Plasmonic Nanopatch Antennas With Integrated Quantum Emitters

*Bryson Krause,[1] Dhananjay Mishra,[2] Jiyang Chen,[1] Christos Argyropoulos[2] and Thang Hoang[1,\*]*

[1]Department of Physics and Material Science, University of Memphis, Memphis, TN 38152
[2]Department of Electrical and Computer Engineering, University of Nebraska-Lincoln, Lincoln, NE, 68588

\*E-mail: tbhoang@memphis.edu

**Optical setup**

In our experimental setup, the NPA sample is characterized by a Nikon LV-150 bright/dark field microscope. The microscope is modified such that an excitation laser can be guided to the sample from above through a set of interchangeable objective lenses, and the radiative SHG signal is collected through the same path in the reflection configuration. The excitation laser wavelength (Coherent Chameleon, 80MHz, 150 fs) is tunable from 680-1080 nm and the SHG signal is dispersed by a spectrometer (Horiba iHR550) and then detected by a charge coupled device (Horiba Synapse). The laser excitation input is spectrally filtered by a 700 nm long-pass filter while the SHG signal is filtered by a 550 nm short-pass filter. For ensemble of NPA measurements, the sample is excited and collected by a 20X (0.4 NA) objective lens while for single NPA measurements, a 100X (0.9 NA) lens is used. Also, for individual NPA measurements a pinhole aperture is placed at a confocal image plane to spatially select signal from single NPAs. Further, the fundamental plasmonic resonant mode of the NPAs is measured through the reflection from the sample due to white light illumination.

**ZnO spacer gap thickness characterization**

Figures S1(a) and (b) show the atomic force microscopy (AFM) images of two ZnO samples which were deposited with 150 and 300 cycles of the Atomic Layer Deposition (ALD) at 200oC. These images were acquired prior to the nanocube deposition. The edge of the ZnO layers was formed by partially covering the Au film substrate by an adhesive tape which can be removed prior to the AFM imaging.

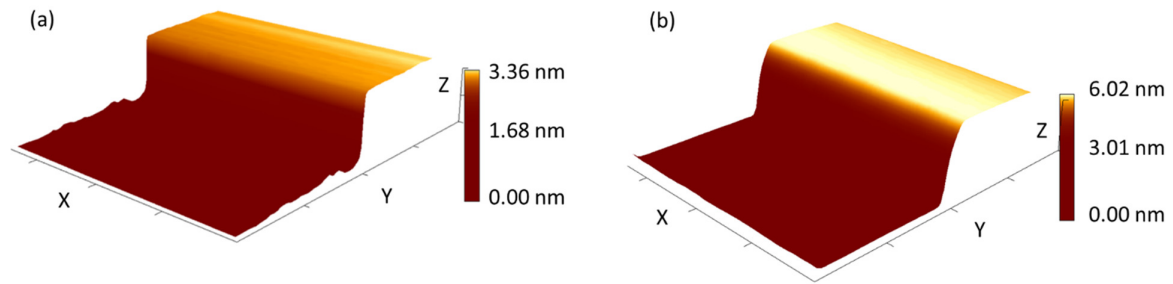

Figure S1. Atomic force microscopy characterizations of ZnO spacer gap thicknesses for two different samples. (a) 150 ALD cycles and (b) 300 ALD cycles.

**Numerical study of SHG response of NPAs**

The numerical analysis of the interaction of light with the NPA, in both linear and nonlinear regimes, was performed in frequency domain using the finite element method (COMSOL Multiphysics software). The structure was illuminated by a plane electromagnetic wave at normal incidence with respect to the z-axis, using continuity periodic conditions on its sides. However, the periodicity was chosen to be large enough in order to emulate a single NPA response. In addition, the simulations were performed in a 2D geometry to reduce the substantial computational burden of nonlinear 3D simulations. To avoid field singularities at the sharp edges of NPA, which are particularly undesirable when modelling nonlinear optical effects, the corners, and the edges of the nanoprisms were rounded with a radius of 15 nm. The top and bottom boundaries of the simulation domain were interfaced by using scattering boundary conditions to ensure the absence of back reflections from the exterior boundary of our computational domain. A 3nm ZnO layer was sandwiched between a 100nm gold substrate layer at the bottom and the 100nm square NPA made of silver at the top.

The numerical study of second harmonic generation was performed by using a two-step procedure in COMSOL Multiphysics. In the first step the interaction of the fundamental incident wave with the nanostructure was modeled by usual linear simulations to determine the local electromagnetic fields at the NPA surface. In the second step, the local distribution of the obtained fundamental field was used to calculate the nonlinear response of the NPA. The second harmonic polarization was computed by using the formula $P_{2\omega}(r) = \chi^{(2)}E^2_\omega(r)$ where $\chi^{(2)}$ is the nonlinear susceptibility matrix of ZnO and $E_\omega$ is the fundamental electric field component inside the ZnO gap layer. The nonlinear components of the second-order nonlinear susceptibility matrix of ZnO used here are taken from the literature[1-2] to be: $d_{15}$ = 10.2 pm/V, $d_{31}$ = 1.36 pm/V and $d_{33}$ = -7.4 pm/V, where $\chi^{(2)}_{ij} = 2d^{(2)}_{ij}$ by convention.[3] Further, we account for the pulse laser excitation in our experiments by multiplying the continuous wave simulation results obtained in the frequency domain with a frequency-domain Gaussian profile that imitates the envelope of the pulsed laser illumination.

**Exciton - plasmon coupling at the fundamental frequency**

Due to the broad resonance band of NPAs, as determined by the white light reflection spectrum, it is very challenging to observe the usual energy splitting spectrum signature of the exciton-plasmon coupling.[4] In order to observe a small energy splitting at the fundamental frequency, we turned to examine the reflection of the pump laser as it is scanned through the fundamental frequency. We observed a broadening and splitting of the laser line from a NPA sample which has a resonance peak at approximately 970 nm as shown in Figure S2. The splitting energy is determined to be 9 meV which is close to $\Delta E/2$ with $\Delta E = 19$ meV, i.e., the splitting at the SHG that was presented in the main text. Compared to the energy splitting observed for the reflected laser at the pump frequency, the splitting of the SHG peak presented in the main text is broader and becomes more obvious and pronounced.

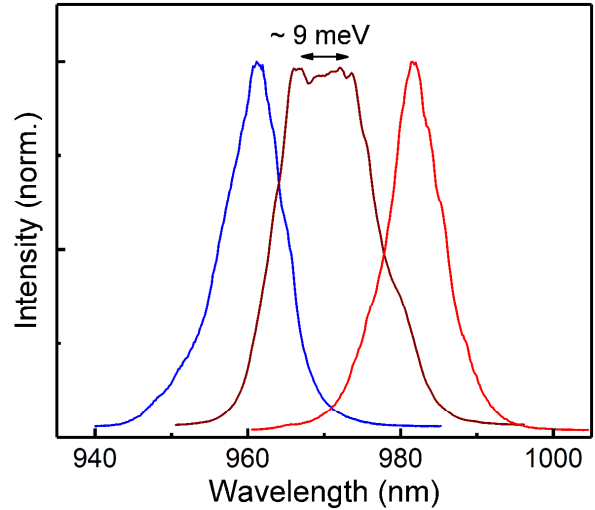

Figure S2. Reflection of pump laser as it is scanned through the fundamental plasmon resonance of NPAs